\documentclass[12pt]{article}

\usepackage{scicite}

\usepackage{times}
\usepackage{graphicx}
\usepackage{float}

\newenvironment{sciabstract}{%
\begin{quote} }
{\end{quote}}

\newcounter{lastnote}

\font\myfont=cmr12 at 15pt

\title{\myfont {ARTIFICIAL INTELLIGENCE ACROSS COMPANY BORDERS} }

\author
{Olga Fink, Torbj{\o}rn Netland, Stefan Feuerriegel\\
\normalsize{ETH Zurich, Switzerland}
}

\date{}

\begin{document} 


\baselineskip24pt


\maketitle


\begin{sciabstract}
Artificial intelligence~(AI) has become a valued technology in many companies. At the same time, a substantial potential for utilizing AI  \emph{across} company borders has remained largely untapped. An inhibiting factor concerns disclosure of data to external parties, which raises legitimate concerns about intellectual property rights, privacy issues, and cybersecurity risks. Combining federated learning with domain adaptation can provide a solution to this problem by enabling effective cross-company AI without data disclosure. In this Viewpoint, we discuss the use, value, and implications of this approach in a cross-company setting.  
\end{sciabstract}

\section*{\fontsize{14}{16} \selectfont The need for artificial intelligence across company borders}

Artificial intelligence (AI) has potential to increase global economic activity in the industrial sector by \$13 trillion by 2030 \cite{McKinsey2018}. However, this potential remains largely untapped because of a lack of access to or a failure to effectively leverage data across companies borders \cite{WEF2020}. AI technologies benefit from large amounts of representative data --  often more data than a single company possesses. It is especially challenging to achieve good AI performance in industrial settings with unexpected events or critical system states that are, by definition, rare. Industrial examples are early detections of outages in power systems or predicting machine faults and remaining useful life, for which robust inference is often precluded. 

A solution is to implement cross-company AI technologies that have access to data from a large cross-company sample. This approach can effectively compile large-scale representative datasets that allow for robust inference. In principle, this could be achieved through data sharing. However, due to confidentiality and risk concerns, many companies remain reluctant to share data directly -- despite the potential benefits \cite{WEF2020}. In some cases, data sharing is also precluded by privacy laws (for example, when involving data from individuals). Likewise, sharing code for AI models among companies has other drawbacks. In particular, it prevents that AI learns from large-scale, cross-company data, and, hence, potential performance gains from cross-company collaboration would be restrained.

To overcome the limitations of direct data sharing, we discuss a potential remedy by using federated learning with domain adaptation. This approach can enable inference across company borders without disclosing the proprietary data. Earlier works discuss the importance of AI in inter-organizational settings (e.g., via meta learning or transfer learning). For instance, in Hirt et al. \cite{hirt2018service}, a prediction ensemble across different inter-organizational entities is built, which is effective when all entities solve the {same} task. What makes federated learning combined with domain adaptation appealing is its flexibility when tasks vary across companies: it allows one to train a collaborative model that is tailored to the specific task and the specific conditions of a company. 



\section*{\fontsize{14}{16} \selectfont Hurdles in collaborating on artificial intelligence}

Two prime hurdles hinder companies from collaborating in AI initiatives. First, a privacy-preserving solution is required so that inference can be made without disclosing the underlying data \cite{WEF2020}. Physical sharing of data could disclose proprietary information on operational processes or other intellectual property to competitors. This is particularly problematic whenever companies seek AI collaboration with suppliers, customers, or competitors. For example, data from manufacturing plants could reveal parameter settings, product compositions, throughput rates, yield, routing, and machine uptimes. If such data are revealed, it can be misused by customers in negotiations or help competitors improve their productivity or products. Besides intellectual property, a number of further constraints are reducing companies' propensity to share data. Examples include trust, cybersecurity risks, ethical constraints, and laws for ensuring a user's right to privacy. 

The second hurdle is that collaborating companies need to account for the possibility of domain shifts. A domain shift refers to discrepancies among the data distributions collected for systems with different configurations or operating conditions \cite{wang2020missing}. For example, machine data from one company may not be representative of operating conditions observed in another company. A domain shift presents hurdles to the underlying inferences: a model that was trained on data from one company is likely to perform poorly when deployed at another company with distinctly different settings or conditions. 


\section*{\fontsize{14}{16} \selectfont Towards artificial intelligence across companies}

\begin{figure}[H]
    \centering
    \includegraphics[scale=1]{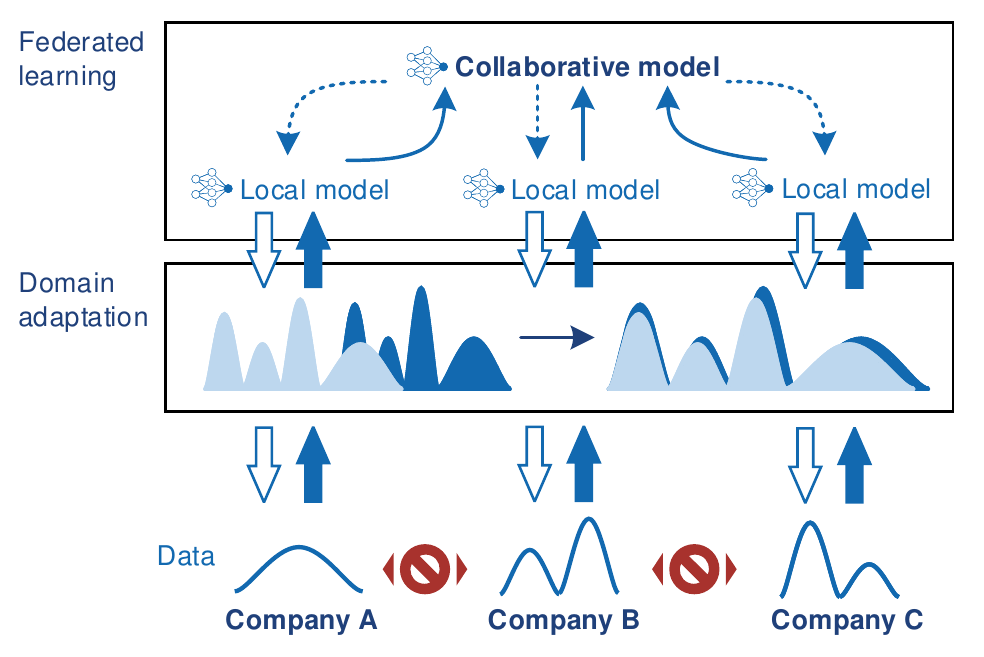}
    \caption{Artificial intelligence across companies by combining federated learning and domain adaptation. Inferences are made through a collaborative model while preserving the privacy of proprietary company data and making the learned representations invariant to the specific operating conditions of the participating companies.
    }
    \label{fig:framework}
\end{figure}

Recent advances in AI research can help overcome these two hurdles. Specifically, we review how cross-company AI can be achieved through a combination of (1)~federated learning to address the privacy-preserving data sharing hurdle and (2)~domain adaptation to address the domain shift hurdle (Figure~\ref{fig:framework}). Such a combination is typically referred to as \emph{federated transfer learning} \cite{Kairouz2019}.
\footnote{Two types of transfer learning approaches can be commonly encountered in industrial ecosystems where faults are considered as labels and are typically rare for safety critical systems: (1)~labels are available in the source domain but not in the target domain (referred to as unsupervised domain adaptation) and (2)~labels are neither available in the source nor in the target domain (referred to as unsupervised transfer learning) \cite{Pan2010}.}



\emph{Federated learning} provides distributed machine learning that preserves privacy \cite{McMahan2017,yang2019federated}. It was mainly introduced for decentralized communication-efficient training on distributed edge devices. In essence, federated learning is a machine learning framework that enables multiple entities to collaborate in solving a machine learning problem while keeping the data from each entity locally, so that only model parameters are shared \cite{Kairouz2019}. Originally, the entities were mobile consumer devices, whereas the entities in our setting are companies. 




In company collaborations, federated learning enables inferences from the joint data of all companies (e.\,g., from several factories, rolling stock, or power plants) without the need to directly share the data. Instead, only model parameters (either gradients or weights) are exchanged between companies. To achieve this, federated learning decouples model training from access to the raw training data: it first trains local models based on the data of each company, then in a subsequent step, the local models are aggregated into a collaborative model that permits population-wide inference. Hence, the collaborative model combines data (and, thus, characteristics from the operational experience) of all collaborating companies, while, during this process, no company has direct access to the training data of other companies. As a result, federated learning in cross-company AI occurs ``vertically'' (Figure~\ref{fig:framework}). In the literature, this specific variant of federated learning is subsumed under \emph{cross-silo} or \emph{feature-based federated learning}.

\vspace{0.3cm}
\emph{Domain adaptation} is required to address the domain shift among data from different companies since the operational data distributions between the companies are typically only partially overlapping. In the context of cross-company AI, its objective will be to make the learned representations invariant to the specific operating conditions of the collaborating companies and, thereby, mitigate the domain shift between the highly varying operating conditions across companies \cite{Pan2010, wang2020missing}. Recent deep domain adaptation approaches learn features that become indistinguishable between the source and target domains, so that their marginal distributions are aligned in the feature space \cite{ganin2015unsupervised}.

\section*{\fontsize{14}{16} \selectfont Bringing cross-company AI into practice}


Companies can combine federated learning and domain adaptation to achieve collaborative AI in industrial ecosystems. Once deployed, it allows cooperating companies to benefit from the collaborative AI model while keeping the raw training data local (and thus concealed). Nevertheless, the collaborative model is trained in a way that it generalizes well across the data from each company. However, at no point are proprietary data shared across company borders. Instead, only models (e.\,g., gradients) are shared across companies. Moreover, the collaborative model accounts for the heterogeneity across companies by making the learned representations invariant to the specific operating conditions of the company -- enabling each participating stakeholder to extend its own operating experience by that of other collaborating companies. 

For industrial ecosystems, we foresee considerable value from implementing cross-company AI in a decentralized manner. Traditionally, the training process in federated learning is typically coordinated by a central player. This may on the one hand impose a potential vulnerability due to the bottleneck character of the setup. On the other hand, this centralized architecture may be less suitable in setups where bilateral collaborations are more prevalent \cite{Kairouz2019}. Fully decentralized learning setups were, therefore, introduced \cite{lalitha2018fully}. In decentralized federated learning, the communication with the central server is then replaced by the peer-to-peer communication. This could be particularly interesting for cross-company collaboration within sub-networks that are determined by the similarity of the applications or the operating conditions. Such sub-networks could then be formed dynamically, depending on the specific use cases and the evolution of the operating conditions. To overcome the need for a central managing body, the use of distributed ledger technology appears promising in this regard. Eventually, the approach discussed here requires learning from empirical implementations, so that companies get guidance on whether to prefer a centralized or decentralized approach.


While federated learning is able to provide significant privacy improvements and may, thereby, encourage collaboration across company borders, it is critical to remind companies that, hitherto, no formal guarantee of privacy can be provided in the standard federated learning model \cite{Kairouz2019}. It might be possible to infer some information from the gradient update and the previous model. Furthermore, standard federated learning techniques can be vulnerable to causative attacks, by which a well-trained model can be corrupted through false model updates \cite{Kairouz2019}. For companies, it is important to choose an implementation where such issues can be obviated. One solution could be to additionally incorporate privacy-preserving technologies such as differential privacy or cryptography.



For practitioners, bringing cross-company AI into industrial ecosystems will require design principles that guide the implementation process. For instance, if there is no significant domain shift between the applications of two companies, federated learning can be applied without combining it with domain adaptation. In addition, implementations of cross-company AI must fulfill further needs from practice (e.g., see the discussion in \cite{hirt2018service}). This may necessitate further extensions such as continuous learning and solutions for data heterogeneity. For instance, for highly heterogeneous systems, an implementation must be chosen that is sufficiently robust and thus, enables transferability (e.\,g., across different product types, different sets of sensors, or different manufacturers). Over time, further work should be channeled into developing consistent standards. 



\section*{\fontsize{14}{16} \selectfont Way forward}

Combining federated learning and domain adaptation can release the power of AI in cross-company settings. Such cross-company AI can be scaled beyond traditional supply chains or domains---creating large ecosystems of collaborating organizations. While this vision may be realized in the near future, companies could start learning this new technology in smaller, trusted partnerships. One reason is that it remains to develop fair value-capture distribution models, which requires a better understanding of the micro-economic implications of cross-company AI.

Aligned with systems thinking, industry managers should identify data partners who can help optimize their performance more holistically than they can on their own. Cross-company AI can also spur new business models, for example through AI-as-a-service orchestrated by third-party companies (e.\,g., \cite{hirt2018service}) Particularly small and medium-sized companies stand to benefit from tapping into data resources of other companies. In this regard, service systems engineering can help develop systematic principles for designing and developing service system networks based on cross-company AI. A first step toward that direction is to acquire a systematic understanding of value co-creation patterns across stakeholders and resources (cf. \cite{bohmann2014service}).

Leveraging AI across companies will benefit from ongoing research. Several attempts are currently underway to advance federated learning, improving its scalability, robustness, and effectiveness while balancing increased privacy protection and model performance. When these issues are sorted, federated learning with domain adaptation can foster an exponential growth in the use of AI across company borders.


\section*{\fontsize{14}{16} \selectfont Acknowledgements} \label{sec:acknowledgements}

The contribition of Olga Fink was funded by the Swiss National Science Foundation (SNSF) Grant no. PP00P2 176878.

\bibliographystyle{unsrt}

\end{document}